\documentstyle[epsf,mnras_cite]{mn}
\title{Estimating non-gaussianity in the microwave background}
\author[A.F. Heavens]{A.F. Heavens\\ 
Institute for Astronomy, University of Edinburgh, 
Blackford Hill, Edinburgh EH9 3HJ, U.K.}

\newcommand{\be}{\begin{equation}}
\newcommand{\ee}{\end{equation}}
\newcommand{\ba}{\begin{eqnarray}}
\newcommand{\ea}{\end{eqnarray}}
\newcommand{\mat}{{\bf}}
\newcommand{\nn}{\nonumber\\}

\begin{document}
\maketitle

\begin{abstract}
The bispectrum of the microwave background sky is a possible discriminator 
between inflationary and defect models of structure formation in the Universe.
The bispectrum, which is the analogue of the temperature 3-point correlation
function in harmonic space, is zero for most inflationary models, but 
non-zero for non-gaussian models.  The expected departures from zero are 
small,  and easily masked by noise, so it is important to be able to estimate
the bispectrum coefficients as accurately as possible, and to know the
errors and correlations between the estimates so they may be used in
combination as a diagnostic to rule out non-gaussian models. 
This paper presents a method for estimating in an unbiased way the 
bispectrum from a microwave background map in the near-gaussian limit.  
The method is optimal, in the 
sense that no other method can have smaller error bars,  and in addition, 
the covariances between the bispectrum estimates are calculated explicitly.
The method deals automatically with partial sky coverage and 
arbitrary noise correlations without modification.
A preliminary application to the Cosmic Background Explorer 4-year dataset
shows no evidence for non-gaussian behaviour.
\end{abstract}

\section{Introduction}

The question of whether or not the microwave background sky is 
well-approximated by a gaussian random field
is important for distinguishing inflationary and defect models 
of the early Universe.    Inflationary models predict immeasurably small 
non-gaussian components, arising from gravity waves \cite{BMS97} and the 
Rees-Sciama effect (\pcite{MGLM95},\pcite{MSS95}).
The difficulty which besets such tests is that the predicted departures
from gaussian behaviour for most defect models are small 
(e.g. \pcite{Luo94}),  
and are correspondingly difficult to detect in the presence of noise, 
which may be 
instrumental or cosmic variance.  This makes it extremely important 
to be able to calculate the statistical properties of the non-gaussian
discriminants.  In particular, it will probably be necessary to combine
the results of a large number of estimates, to obtain a statistically 
significant departure from gaussianity (if it exists), or to make a
convincing case that the sky is indeed gaussian.   

Ironically, the best evidence
for an inflationary model may well come not from a specific test for
non-gaussianity, but rather from the power spectrum.   For inflationary 
models, the power spectrum of temperature fluctuations is predicted to 
have a reasonable amount of 
structure in it, with multiple acoustic oscillation peaks which should be 
measurable by future satellites such as MAP and Planck (\pcite{Jungman}, 
\pcite{PhaseA}).  
Should such structures be found, and found to agree with the inflationary
model predictions within the errors, the case against any
form of defect model would be strong, and even the present knowledge 
of the power spectrum appears already to rule out many defect models
(\pcite{PST97}).   In this case, the absence of non-gaussian signatures 
would be a
useful confirmation.  However, if the power spectrum turns out not to be 
well-fitted by inflationary models, the question of the gaussian nature or 
otherwise of the fluctuations becomes correspondingly more important.
There are many ways to approach the problem of determining whether 
fluctuations are gaussian, in large-scale structure and in the CMB.  
These include the 3-point function 
(e.g. \pcite{Hinshaw94}, \pcite{FRS93}, \pcite{LS93}, \pcite{GLMM94}), 
the genus and Euler-Poincar\'e statistic (\pcite{Coles88}, \pcite{Gott90},
\pcite{Luo94b}, \pcite{Smoot94}),  peak statistics (\pcite{BE87}, 
\pcite{Kogut95}, \pcite{Kogut96}) 
and studies of tensor modes in the CMB \cite{CCT94}.
The approach we take here is to investigate the bispectrum, for which some 
studies have been made in the large-scale structure literature (\pcite{HBCJ95},
\pcite{MVH97}, \pcite{VHM98}) and in the CMB \cite{Luo94}.   There may be
sharper tools for detecting {\em specific} non-gaussian models, but the 
rationale 
for this approach is that the bispectrum offers a {\em generic} test for 
non-gaussian models, in the following sense:  a general field will 
have non-zero connected $n$-point functions at all
orders,  and the bispectrum is the lowest statistic (with $n>1$) for which a 
gaussian field has zero expectation value.  

The principal advantages of the approach detailed in this paper are 
that it deals automatically with masked regions of the sky and 
correlated noise, 
and that the estimates of the bispectrum coefficients come with error bars
and covariances between the errors.   This last point is particularly 
important when one realises that a single bispectrum coefficient estimate
is unlikely to rule out a model, because the cosmic variance is often
larger than the expected signal, and so one is going to need many coefficients
in practice.  A final point is that, for gaussian fluctuations, the estimator
below cannot be improved, in the sense of having a smaller error bar.

\section{Method}

The optimisation procedure in this paper is a generalisation of the optimal
quadratic estimator for the power spectrum, presented by \scite{Teg97a}.  
For consistency, we follow his notation as far as possible.
Let $x_i$ be the temperature fluctuation $\Delta T/T$, in some sky pixel $i$.
The temperature map is expanded in spherical harmonics $Y_\ell^m$ in the 
usual way:
\be
a_{\ell m} \equiv \int\,{\rm d}\Omega\, {\Delta T\over T} 
Y_\ell^{m*}(\Omega) \simeq \sum_i x_i Y_\ell^{m*}(\theta_i, \phi_i)
\Delta\Omega_i
\ee
where ${\rm d}\Omega$, $\Delta\Omega_i$ represent elements of solid angle, 
and $\theta$ and
$\phi$ are polar coordinates.  The power spectrum is defined as
\be
C_\ell = \langle |a_{\ell m}|^2\rangle
\ee
where the angle brackets indicate an ensemble average.  Expectation values of
products of distinct spherical harmonic coefficients are zero by isotropy,
independently of whether the temperature map is gaussian or not.
The bispectrum is defined as
\be
B(\ell_1\ell_2\ell_3,m_1 m_2 m_3) 
\equiv \langle a_{\ell_1}^{m_1} a_{\ell_2}^{m_2} a_{\ell_3}^{m_3} \rangle.
\ee
It is zero, unless the indices comply with the following triangle closure 
constraints (e.g. \pcite{Edmonds}, \pcite{Luo94}):
$m_1+m_2+m_3=0$; $\ell_1+\ell_2+\ell_3=$even; $|\ell_i-\ell_j|\le \ell_k \le
\ell_i+\ell_j$ for $i,j,k=1,2,3$. 

We seek an estimator of $B$ 
which is lossless, if possible, in the sense that it contains as much
information as the original map $\{x_i\}$.  
Ideally it should be unbiased, and with
calculable statistical properties.    In the spirit of Tegmark's optimal
quadratic estimator for the power spectrum,  we seek an estimator for
the bispectrum which is cubic.  We consider quantities 
$y_\alpha$ of the following form
\be
y_\alpha = \sum_{{\rm pixels}\ ijk} E^\alpha_{ijk}\ x_i x_j x_k.
\ee
We will find that the $y_\alpha$  are related to the bispectrum estimates, 
but will not be the bispectrum estimates themselves.
We introduce the shorthand notation $\alpha \equiv 
\{\ell_1,\ell_2,\ell_3,m_1,m_2,m_3\}$, and we also combine the list of 
data triplets into a data vector with elements labelled by $A$: 
\be
D_A \equiv x_i x_j x_k
\ee
where $A$ represents some triplet $\{i,j,k\}$.  The $E^\alpha_{ijk}$ are some
coefficients to be determined.  The mean of $y_\alpha$ involves the 
3-point function, which may be written in terms of the bispectra as follows:
\be
\mu_A\equiv \langle x_i x_j x_k\rangle = \sum_\alpha  B_\alpha R_{ijk}^\alpha
\ee
where we have assumed that the noise has a zero 3-point function.  If it is
known and non-zero, it may be added.  The functions connecting the 3-point
functions in real and harmonic space are \cite{GLMM94}:
\begin{eqnarray}
R_{ijk}^\alpha & = & {\pi\over 2}N(\gamma_{ij},\gamma_{jk},\gamma_{ki}) 
W_{\ell_1}W_{\ell_2}
W_{\ell_3} \nn
&\times & \sum_{{{\ell_4 \ell_5 \ell_6}\atop {m_4 m_5 m_6}}} P_{\ell_4} 
(\cos\gamma_{ij})
P_{\ell_5} (\cos\gamma_{jk}) P_{\ell_6} (\cos\gamma_{ki})\nn
& \times & H_{\ell_4 \ell_6 \ell_1}^{m_4 m_6 m_1} H_{\ell_5 \ell_4 
\ell_2}^{m_5 m_4 m_2} H_{\ell_5 \ell_6 \ell_3}^{m_5 m_6 m_3}\nn
\end{eqnarray}
where 
\be
H_{\ell_1 \ell_2 \ell_3}^{m_1 m_2 m_3} = \int d\Omega\, Y_{\ell_1}^{m_1*} 
Y_{\ell_2}^{m_2}Y_{\ell_3}^{m_3}
\ee
and can be related to Clebsch-Gordon coefficients.   The effect of 
beam-smearing
(here modelled by a gaussian) is through the window functions
\be
W_\ell = \exp\left[-\ell(\ell+1)
\sigma^2/2\right].
\ee
$\gamma_{ij}$ is the angle between pixels $i$ and $j$, and
\ba
N^2(\gamma_{ij},
\gamma_{jk},\gamma_{ki}) & \equiv & 1-\cos^2\gamma_{ij}-\cos^2\gamma_{jk}-
\cos^2\gamma_{ik}+\nn
& & 2\cos\gamma_{ij}\cos\gamma_{jk}\cos\gamma_{ik}.
\ea

\subsection{Optimal estimator $y_\alpha$}

We wish to minimise the variance of $y$ (cf \pcite{Teg97a} for power spectrum),
which involves the 6-point function.  
The means are 
\be
\langle y_\alpha \rangle = \sum_{\alpha'ijk} 
B_{\alpha'} R^{\alpha'}_{ijk} E^{\alpha}_{ijk}.
\ee
The covariance beween the $y$s is
$C_{\alpha\alpha'}\equiv \langle y_\alpha y_{\alpha'}\rangle - 
\langle y_\alpha\rangle\langle y_{\alpha'}\rangle$ which we obtain from the
triplet data covariance matrix:
\be
\langle x_i x_j x_k x_{i'} x_{j'} x_{k'}\rangle - 
\langle x_i x_j x_k\rangle\langle x_{i'} x_{j'} x_{k'}\rangle.
\ee
We now make an assumption concerning the departures from gaussianity.  Since
these are expected to be small, we approximate the covariance matrix by the
covariance matrix for a gaussian field with the same power spectrum.  This
assumes that the bispectrum is small compared with the cosmic variance,
and also assumes that the connected 4-point function is small.    Strictly,
this method is optimal for testing the hypothesis that the field is gaussian,
but it should be very close to optimal for practical cases, since the 
expectation is that the bispectrum will be small.   If the
assumption is not justified, and the bispectrum is not small compared with the
cosmic variance,  detection will not be difficult in any 
event.  If this turns out to be the case, it will be important to check 
that the estimator remains unbiased in the case of large intrinsic bispectrum.

In the gaussian approximation, $\langle x_i x_j x_k\rangle=0$, and we use 
Wick's theorem to write
\begin{eqnarray}
\langle x_i x_j x_k x_{i'} x_{j'} x_{k'}\rangle &  = & 
\xi_{ij}\xi_{ki'}
\xi_{j'k'}
\nn
& & {\rm\ +  \ permutations\ (15\ terms).}
\end{eqnarray}
where we have defined the 2-point function of the temperature field:
\be
\xi_{ij} \equiv \langle x_i x_j\rangle = \sum_\ell {2\ell+1\over 4\pi} C_\ell
P_\ell (\cos\gamma_{ij})W_\ell^2 + N_{ij}
\ee
$P_\ell$ are Legendre polynomials and $N_{ij}$ is the noise 
covariance matrix.
We can then compute the covariance matrix for $y_\alpha$:
\be
V_{\alpha\alpha'} \equiv \langle y_\alpha y_{\alpha'} \rangle 
=  \left(\xi_{ij}\xi_{ki'}
\xi_{j'k'} {\rm\ +  \ perm.}\right) E^{\alpha}_{ijk} E^{\alpha'}_{i'j'k'}
\ee
and we have from now adopted the summation convention for repeated pixel
indices, and also, unless stated otherwise, $\alpha$ indices.
The products of $\xi$ terms are of two types:  there are 6 terms like
$\xi_{ii'}\xi_{jj'}\xi_{kk'}$, with one of each pair of subscripts 
from each distinct $E$, and 9 terms of the form $\xi_{ij}\xi_{ki'}\xi_{j'k'}$
where only one $\xi$ mixes dashed and undashed indices.  
Since the $E^{\alpha}_{ijk}$ are symmetric
to permutations in the $\{ijk\}$, we get
\be
V_{\alpha\alpha'} =  \xi_{ii'}\left(6 \xi_{jj'}\xi_{kk'}+
9 \xi_{jk}\xi_{j'k'}\right)E^{\alpha}_{ijk} E^{\alpha'}_{i'j'k'}.
\label{V}
\ee
We now minimise the variance $V_{\alpha\alpha}$ (not summed) with respect to
$E^{\alpha}_{ijk}$, subject to
a normalisation constraint on $E$ to ensure it is not driven to zero.  We
choose
\be
R^{\alpha}_{ijk} E^{\alpha}_{ijk}=1,
\ee
giving
\be
  \xi_{ii'}\left(6 \xi_{jj'}\xi_{kk'}+
9 \xi_{jk}\xi_{j'k'}\right)E^{\alpha}_{i'j'k'} = 
\lambda R^{\alpha}_{ijk}
\ee
where $\lambda$ is a Lagrange multiplier.  Multiplying by 
$\xi^{-1}_{j''j}\xi^{-1}_{i''i}$, and summing over $ij$, gives
\be
6 \xi_{kk'} E^\alpha_{i''j''k'}+ 9 \xi_{j'k'} \delta^K_{j''k}
E^{\alpha}_{i''j'k'}=
\lambda \xi^{-1}_{j''j}\xi^{-1}_{i''i} R^{\alpha}_{ijk}.
\ee
$\delta^K_{ij}$ is a Kronecker delta function.
Defining $r^\alpha_{i''jk}=\xi^{-1}_{i''i}R_{ijk}^\alpha$, we get,
after some relabelling of indices,
\be
6 \xi_{ij} E^\alpha_{i''jk}+ 9\delta^K_{ik}
\xi_{jm}E^\alpha_{i''mj}=\lambda \xi^{-1}_{ij}r^\alpha_{i''kj}.
\ee
Taking the trace of this equation, and inserting it into the second term
gives the coefficients we require for $y_\alpha$ to have its minimum 
error bar:  
\be
E^\alpha_{ijk} = {1\over 6}\xi_{ii'}^{-1}\left[\xi_{jj'}^{-1}\xi_{kk'}^{-1}
 - {3\over 2+3N}\,\xi_{jk}^{-1}\xi_{j'k'}^{-1}\right]
R_{i'j'k'}^\alpha.
\label{E}
\ee
In this expression, $\lambda$ has been set to unity for convenience,
and $N$ is the number of pixels in the map.
Provided that $\lambda$ is finite, its value does not affect the 
bispectrum information content of $y_\alpha$:  any multiple of $y_\alpha$  
contains the same information as $y_\alpha$ itself.  This makes obvious 
sense, and can be shown formally via the Fisher information matrix (below).

\section{Lossless Estimator of the Bispectrum}  

In order to estimate how well the $y_\alpha$ will perform in estimating
the desired parameters $B_\alpha$, we compute the Fisher information
matrix \cite{TTH97}
\be
F_{\alpha \alpha'} \equiv -\langle {\partial^2 \ln p\over \partial B_{\alpha}
B_{\alpha'}}\rangle
\ee
where $p$ is the posterior probability distribution for the parameters 
(equal to the likelihood, if uniform priors for the parameters are assigned). 
For a data vector with components with means $\bf{\mu}$ and covariance matrix 
${\mat C}$ the Fisher matrix is
\be
F_{\alpha \alpha'} = {1\over 2}{\rm Tr}\left[C^{-1}{\partial C \over 
\partial B_\alpha} C^{-1}{\partial C \over \partial B_{\alpha'}} + 2C^{-1}
{\partial \mu \over \partial B_\alpha}{\partial \mu \over \partial B_{\alpha'}}
\right].
\ee
The error on the parameters $B_\alpha$ is contained in this matrix:  
if all other parameters are known, the minimum error is the conditional one, 
$\sigma_{B_\alpha} = 1/\sqrt{F_{\alpha\alpha}}$.   If all parameters are to
be estimated from the data, then the appropriate error is the marginal 
error $\sqrt{F^{-1}_{\alpha\alpha}}$.  This assumes the probability surface
is adequately approximated by a second-order Taylor expansion at the peak.

As expected for the `near-gaussian' approximation, the covariance matrix 
for either the triplets $x_i x_j x_k$ or the $y_\alpha$ does not 
depend on the parameters to be estimated -- the Fisher matrix is determined
only by the derivatives of the mean values.   For the triplets,
\begin{eqnarray}
F_{\alpha \alpha'} & = &{\mat C}_{ijki'j'k'}^{-1} R_{ijk}^{\alpha} R_{i'j'k'}^
{\alpha'}\nn
& = & \left[\xi_{ii'}\left(6\xi_{jj'}\xi_{kk'}+9\xi_{jk}\xi_{j'k'}\right)
\right]^{-1}R_{ijk}^{\alpha} R_{i'j'k'}^{\alpha'}
\label{Ftrip}
\end{eqnarray}
A similar procedure to the computation of $E$ above gives the inverse 
covariance matrix
\be
{\mat C}_{ijki'j'k'}^{-1}={1\over 6}\xi^{-1}_{kk'}\xi^{-1}_{jj'}\xi^{-1}_{ii'}
-{1\over 2(2+3N)}
\xi^{-1}_{j'k'}\xi^{-1}_{jk}\xi^{-1}_{ii'}
\label{invC}
\ee
This is an important simplification for computational reasons:  without it,
the inversion of an $N^3 \times N^3$ matrix ($C$) would be impractically
slow.  Decomposing its inverse into $N \times N$ matrices $\xi^{-1}$ is much
faster.  
 
Since we can recreate the original temperature map $\{x_i\}$ from the triplets,
(\ref{Ftrip}) is also the Fisher information matrix for the original,
entire map.
We now make a comparison with the Fisher matrix for the $y_\alpha$ -- are
their errors as small as is possible with the entire map?
The covariance matrix for the $y_\alpha$ (\ref{V}) is, for the optimal
choice of $E$ coefficients (\ref{E}),
\ba
V_{\alpha\alpha'} & = &  \xi_{ii'}\left(6 \xi_{jj'}\xi_{kk'}+
9 \xi_{jk}\xi_{j'k'}\right)\nn
& \times & 
C^{-1}_{ijki''j''k''}
R_{i''j''k''}^\alpha 
C^{-1}_{i'j'k'i'''j'''k'''}
R_{i'''j'''k'''}^\alpha\nn
& = & F_{\alpha\alpha'}
\ea

Since the ensemble average of $y_\alpha$ is 
\begin{eqnarray}
\langle y_\alpha \rangle & = & 
B_{\alpha'} R^{\alpha'}_{ijk} E^{\alpha}_{ijk}\nn
& = & B_{\alpha'} R^{\alpha}_{ijk} R^{\alpha'}_{i'j'k'}
C^{-1}_{ijki'j'k'}
\end{eqnarray}
i.e. we obtain the simple result
\be
\langle y_\alpha \rangle = B_{\alpha'}F_{\alpha\alpha'}.
\ee
Consequently, we can use the vector
\be
\hat B_\alpha = F_{\alpha \alpha'}^{-1}y_{\alpha'}
\ee
as an estimator of the bispectrum.   It is unbiased:
\begin{eqnarray}
\langle \hat B_\alpha\rangle & = & F_{\alpha \alpha'}^{-1}\langle y_{\alpha'}
\rangle\nn
& = & F_{\alpha \alpha'}^{-1} B_{\alpha''}F_{\alpha'\alpha''} = B_\alpha.
\end{eqnarray}
and the bispectrum estimates also have calculable covariance properties:
\begin{eqnarray}
C_{\alpha\alpha'} & \equiv & \langle \hat B_\alpha \hat B_{\alpha'}\rangle-
\langle \hat B_\alpha\rangle
\langle \hat B_{\alpha'}\rangle\nn 
& = &  F_{\alpha\alpha''}^{-1} F_{\alpha'\alpha'''}^{-1} \langle y_{\alpha''}
 y_{\alpha'''}\rangle - F_{\alpha \alpha''}^{-1} 
F_{\alpha' \alpha'''}^{-1} \langle y_{\alpha''} \rangle \langle y_{\alpha'''}
\rangle\nn
& = & F_{\alpha\alpha''}^{-1} F_{\alpha'\alpha'''}^{-1}  F_{\alpha''\alpha'''}
\nn
& = & F_{\alpha\alpha'}^{-1}.
\end{eqnarray}
This also proves that the estimators are optimal, by the Fisher-Cramer-Rao
inequality.

\section{Application to COBE 4-year DMR data:}

We illustrate the method by applying the method to the COBE DMR 4-year data, 
focussing on measuring low-order coefficients. For this
experiment, the width of the approximately gaussian beam is $\sigma=3.2^\circ$
\cite{Wright92}.  The method is computationally expensive, and is in the
process of being optimised, but for the moment, the approach taken is to 
average the $\sim 4000$ unmasked pixels of the COBE dataset into larger pixels
of roughly 12 degrees square.  This introduces an additional effective 
gaussian smoothing for large scale coefficients of $12^\circ/\sqrt{12}$,
which is added in quadrature to the COBE beam.  We shall see the effect of
this additional pixelisation below, in the form of an error bar larger than
that of cosmic variance, especially for the higher harmonics.
The effective beam suppresses contributions
to the bispectrum from harmonics with $\ell$ larger than 
$\ell(\ell+1)\sigma_{\rm eff}^2/2 \simeq 1$, i.e. $\ell > 17$.  We therefore
truncate the summations in the estimator for $B$, and the Fisher matrix 
(scalar in this case) at the conservative limit of $\ell_{\rm max}=40$.
Pixel errors are taken from the COBE DMR datasets, and assumed to be 
independent.  Averaging is done by inverse-variance weighting.  The power
spectrum is assumed to have a Harrison-Zeldovich spectrum, and the 
normalisation is $Q_{\rm rms} = 18.4 \mu K$ \cite{GBBHKSW}.
Coefficients are chosen for which non-Gaussian predictions are quoted
in \cite{Luo94}
\begin{eqnarray} 
B(2\,2\,2,1\,1\,-2) & = & (2.5 \pm 4.1) \times 10^{-15} \nn
B(4\,4\,4,2\,2\,-4) & = & (5.4 \pm 6.5) \times 10^{-15}\nn
B(6\,6\,6,3\,3\,-6) & = & (12.4 \pm 8.9) \times 10^{-15}\nn
B(8\,8\,8,4\,4\,-8) & = & (1.5 \pm 11.6) \times 10^{-15}\nn
B(10\, 10\, 10, 5\, 5\, 10) & = & (-16.3 \pm 16.6) \times 10^{-15}\nn
\end{eqnarray}
from which we see that there is no evidence of non-gaussianity, at least 
for these coefficients.  Note that for the first bispectrum estimate, the 
cosmic variance corresponds to an error of $1.4 \times 10^{-15}$ (e.g.
\pcite{Luo94}).  
 These very large-scale modes are not ideal for this sort of study -- a
higher-resolution experiment generally has higher signal-to-noise
\cite{Luo94}.  A recent preprint \cite{Ferreira98} claims a detection
of non-gaussianity at $\ell=16$, a mode which is not probed here.

\section{Summary}

In this paper, we see that it is possible to construct an estimator for 
the bispectrum which is lossless, in the sense that it contains as much
information on the bispectrum as the entire map.  It is also unbiased,
and the covariance properties of the estimators are calculable.  The
estimator involves one approximation -- that the departures from 
gaussianity are small.  In this limit, there is no other method which
will lead to smaller error bars.  The fact that the covariance properties 
are known is important in practical cases, because the bispectrum may well
be small in comparison to cosmic variance, so a single estimate is unlikely
to be sufficient to rule out many non-gaussian models.  Many estimates (with
different $\ell_1\ell_2\ell_3 m_1 m_2 m_3$) would be required.   As a test
for non-gaussianity, this method has the advantage that confidence levels
can be computed analytically, without recourse to Monte Carlo simulation.

The notion of an optimal method is defined rather precisely in terms of
information content and bias, but the issue of whether a method is good
or not is wider than this.   There is no doubt that the number of computations
required to do this analysis is very large, dominated for reasonable 
pixel counts by computation of the $R$ coefficients.  This can be aided by
precomputing Clebsch-Gordan coefficients \cite{CG}
and using a packed-storage algorithm,
and by using parallel computers, for which this problem is ideal. 
However, it is clear that it will not be possible
to deal directly with the entire dataset from Planck without some form
of pre-compression, perhaps along the lines above for the low-$\ell$
modes.  For high-$\ell$ modes, subdivision of the sky into essentially
independent datasets may be required.  This is an inherent problem for
high-order statistics, but it may also be required even for the power 
spectrum.

A further point to note is that the method deals automatically with sky
coverage which is not complete, and arbitrary noise correlations.  
These are the two crucial tests of any method, since they will be a 
feature of future high-quality experiments such as Planck and MAP. 
Three-point function estimators have traditionally not computed the
covariances directly, but rely on simulation tests to decide significance
(e.g. \pcite{Kogut96}).  This is the disadvantage of many methods (e.g. genus,
extrema correlation functions) for which the evaluation of errors may be 
difficult analytically.  Against this one has, of course, to balance speed 
advantages.

A preliminary application to the COBE 4-year data shows 
large-scale bispectrum coefficients consistent with zero.  However, the 
errors are sufficiently large that these coefficients do not rule out 
non-gaussian models with confidence.

\bibliographystyle{mnras}
\bibliography{general}

\end{document}